\documentclass[a4paper,reqno,twoside]{amsart}

\usepackage{latexsym,amssymb}
\usepackage[initials,alphabetic,nobysame]{amsrefs}

\theoremstyle{plain}
\newtheorem{propn}{Proposition}[section]
\newtheorem{thm}[propn]{Theorem}
\newtheorem{lemma}[propn]{Lemma}
\newtheorem{cor}[propn]{Corollary}
\newtheorem*{thm*}{Theorem}

\theoremstyle{definition}

\theoremstyle{remark}
\newtheorem*{rem}{Remark}
\newtheorem*{rems}{Remarks}
\newtheorem*{eg}{Example}
\newtheorem*{egs}{Examples}

\newcommand{\Hil}{\mathsf{H}}
\newcommand{\hil}{\mathsf{h}}
\newcommand{\Kil}{\mathsf{K}}

\newcommand{\Fock}{\mathcal{F}}
\newcommand{\Exp}{\mathcal{E}}
\newcommand{\e}[1]{\varepsilon (#1)}

\newcommand{\step}{\mathbb{S}}

\newcommand{\init}{\mathfrak{h}}
\newcommand{\noise}{\mathsf{k}}
\newcommand{\khat}{{\wh{\noise}}}

\newcommand{\Qcd}{Q^{c,d}}

\newcommand{\ind}{\mathbf{1}}

\newcommand{\Mat}{\mathrm{M}}

\newcommand{\Fh}{F_\frac{1}{2}}
\newcommand{\fh}{f_\frac{1}{2}}
\newcommand{\gh}{g_\frac{1}{2}}

\newcommand{\Alg}{\mathcal{A}}
\newcommand{\Clg}{\mathcal{C}}
\newcommand{\vM}{\mathcal{M}}
\newcommand{\vN}{\mathcal{N}}
\newcommand{\Vop}{\mathsf{V}}

\newcommand{\XF}{X^F}
\newcommand{\FX}{{}^F \! X}

\newcommand{\al}{\alpha}
\newcommand{\be}{\beta}

\newcommand{\Ga}{\Gamma}
\newcommand{\si}{\sigma}

\newcommand{\Real}{\mathbb{R}}
\newcommand{\Rplus}{\Real_+}
\newcommand{\Comp}{\mathbb{C}}
\newcommand{\Nat}{\mathbb{N}}
\newcommand{\Int}{\mathbb{Z}}

\newcommand{\ip}[2]{\langle #1, #2 \rangle}

\newcommand{\norm}[1]{\lVert #1 \rVert}
\newenvironment{sbmatrix}
{\bigl[\begin{smallmatrix}}{\end{smallmatrix}\bigr]}

\newcommand{\wh}{\widehat}
\newcommand{\wt}{\widetilde}
\newcommand{\ol}{\overline}

\newcommand{\ot}{\otimes}
\newcommand{\op}{\oplus}
\newcommand{\uot}{\underline{\otimes}}

\newcommand{\To}{\rightarrow}
\newcommand{\Tends}{\rightarrow}
\newcommand{\Implies}{\ensuremath{\Rightarrow}}
\newcommand{\Iff}{\ensuremath{\Leftrightarrow}}
\newcommand{\ges}{\geqslant}
\newcommand{\les}{\leqslant}

\newcommand{\bd}{{\text{\tu{b}}}}

\newcommand{\ds}{\displaystyle}
\newcommand{\ti}{\textit}
\newcommand{\tu}{\textup}
\newcommand{\ttu}[1]{\text{\tu{#1}}}
\newcommand{\dfn}{\ti}

\DeclareMathOperator{\Ran}{Ran}
\DeclareMathOperator{\Lin}{Lin}

\newenvironment{alist}
{

\begin{enumerate}}
{\end{enumerate}}

\newenvironment{rlist}
{

\begin{enumerate}}
{\end{enumerate}}

\newcounter{step_count}

\numberwithin{equation}{section}

\begin{document}

\title[Generators of operator cocycles]{On The Generators Of Quantum
Stochastic Operator Cocycles}
\author{Stephen J.~Wills}
\address{School of Mathematical Sciences \\ University College Cork \\
Cork \\ Ireland}
\email{s.wills@ucc.ie}
\thanks{This work was begun whilst the author was at the University of
Nottingham, supported by a Lloyd's of London Tercentenary Foundation
Fellowship.}
\dedicatory{Dedicated to the memory of John Lewis --- an inspirational
academic great-grandfather.}

\begin{abstract}
The stochastic generators of Markov-regular operator cocycles on
symmetric Fock space are studied in a variety of cases: positive
cocycles, projection cocycles, and partially isometric
cocycles. Moreover a class of transformations of positive contraction
cocycles is exhibited which leads to a polar decomposition result.
\end{abstract}

\maketitle

\section{Introduction}

Let $\Hil$ be a Hilbert space and $\si = (\si_t)_{t \ges 0}$ be a
semigroup of unital endomorphisms of the algebra $B(\Hil)$ --- an
$E_0$-semigroup (\cite{wAEbook}). A family of operators $X = (X_t)_{t
\ges 0} \subset B(\Hil)$ is a \dfn{left cocycle} (respectively a
\dfn{right cocycle}) for $\si$ if it satisfies
\begin{equation} \label{cocycles}
X_0 = I \ \text{ and } \ X_{r+t} = X_r \si_r (X_t) \quad
(\text{resp.\ } X_{r+t} = \si_r (X_t) X_r)
\end{equation}
for all $r,t \ges 0$. \ti{Unitary} cocycles (i.e.\ each $X_t$ is
unitary) play a fundamental role in the classification of
$E_0$-semigroups, which is carried out up to conjugation by such
objects. The $E_0$-semigroups of type I (those that possess a
sufficiently large number of cocycles) turn out to be precisely those
that are cocycle conjugate to the CCR flow on symmetric Fock space,
the particular flow being uniquely specified by a choice of Hilbert
space $\noise$, called the \dfn{noise dimension space} in the language
of quantum stochastic calculus (QSC). In QSC cocycles arise naturally
as solutions of a quantum stochastic differential equation (QSDE) of
Hudson-Parthasarathy type, and it is standard practice to ampliate the
CCR flow so that it acts on $B(\init \ot \Fock_+)$, where $\init$ is
another Hilbert space (the \dfn{initial space}) and $\Fock_+$ denotes
the symmetric Fock space over $L^2 (\Rplus; \noise)$. The coefficient
driving the QSDE is some operator $F \in B(\init \ot \khat)$, where
$\khat := \Comp \op \noise$, and (conjugation by) the resulting
cocycle can be viewed as a Feynman-Kac perturbation of the free
evolution given by $\si$ (\cite{lAqFK}). Conversely, any contraction
cocycle that is \ti{Markov-regular} necessarily satisfies such a QSDE
for some such $F$, and moreover the collection of those $F$ that are
generators of contraction cocycles is now well-known, as are the
subsets corresponding to the generators of isometric, coisometric and
unitary cocycles. For more details see \cites{fFqp8, LPgran, mother,
father}, or the lecture notes~\cite{jmLGreif}.

In this paper we characterise the generators of many other classes of
cocycle, namely self-adjoint cocycles, positive cocycles, projection
cocycles and partially isometric cocycles, going beyond the case of
contractive cocycles for the first two classes. Positive contraction
cocycles have appeared in the work of Bhat~(\cite{bvrBCCR}), where
they are used to study dilations and compressions between
$E$-semigroups and quantum Markov semigroups on $B(\Hil)$. In the
third section of the paper we discuss a one-parameter family of
transformations on the class of positive contraction cocycles, and
describe the corresponding transformation on the stochastic generators
in the Markov-regular case. This leads naturally to a polar
decomposition result in the final section, where it is shown that any
Markov-regular contraction cocycle with commutative component von
Neumann algebra can be written as a product of a partial isometry
cocycle and a positive cocycle.

\subsection{Notational conventions}

Algebraic tensor products are denoted by $\uot$, with $\ot$ reserved
for the (completed) tensor product of Hilbert spaces and the tensor
product of von Neumann algebras. The tensor symbol between Hilbert
space vectors in elementary tensors will usually be suppressed. Given
Hilbert spaces $\Hil$ and $\hil$, and $x \in \hil$, we define maps $E_x 
\in B(\Hil; \Hil \ot \hil)$ and $E^x \in B(\Hil \ot \hil; \Hil)$ by
\[
E_x = u \ot x = ux \quad \text{ and } \quad E^x = (E_x)^*,
\]
with context indicating the choice of $\Hil$ and $\hil$.

\section{Operator cocycles on Fock space}

Fix two Hilbert spaces, the initial space $\init$ and the noise
dimension space $\noise$. Let $\Fock_+$ denote the symmetric Fock
space over $L^2 (\Rplus; \noise)$, and in general let $\Fock_I$ denote
the symmetric Fock space over $L^2 (I; \noise)$ for $I \subset \Real$.
We shall make frequent use of the time shift and time reversal
operators on $\init \ot \Fock_+$, $S_t$ and $R_t$ respectively, which
are the ampliated second quantisations of
\[
(s_t f)(u) = \begin{cases} 0 & \text{if } u < t, \\ f(u-t) & \text{if
} u \ges t, \end{cases}
\ \text{ and } \ 
(r_t f)(u) = \begin{cases} f(t-u) & \text{if } u \les t, \\ f(u) &
\text{if } u > t, \end{cases}
\]
so that $S_t u\e{f} = u\e{s_t f}$, where $\e{f} = (1, f, (2!)^{-1/2} f
\ot f, \ldots)$ is the exponential vector associated to $f \in L^2
(\Rplus; \noise)$. Note that the $S_t$ are isometries and the $R_t$
are self-adjoint unitaries, with both maps $t \mapsto S_t$ and $t
\mapsto R_t$ continuous in the strong operator topology, that is,
\dfn{strongly continuous}. The endomorphism semigroup $(\si_t)_{t \ges
0}$ on $B(\init \ot \Fock_+)$ is constructed from $(S_t)_{t \ges 0}$
by using the obvious isomorphism these maps induce between $\Fock_+$
and $\Fock_{[t,\infty[}$. More concretely, for any $X \in B(\init \ot
\Fock_+)$ the operator $\si_t (X)$ is determined by
\begin{equation} \label{sig defn}
\ip{u\e{f}}{\si_t (X) v\e{g}} = \ip{u\e{s^*_t f}}{X v\e{s^*_t g}} \exp
\int^t_0 \ip{f(u)}{g(u)} \, du.
\end{equation}
The time reversal operators on $B(\init \ot \Fock_+)$ are
\begin{equation} \label{rho defn}
\rho_t (X) = R_t X R_t.
\end{equation}

A \dfn{Fock-adapted} left (respectively right) cocycle on $\init \ot
\Fock_+$ is any family $X = (X_t)_{t \ges 0} \subset B(\init \ot
\Fock_+)$ that satisfies the functional equation~\eqref{cocycles}
together with the adaptedness condition
\begin{equation} \label{adapted}
X_t \in B(\init \ot \Fock_{[0,t[}) \ot I_{\Fock_{[t,\infty[}},
\end{equation}
where we utilise the continuous tensor product factorisation of Fock
space: $\Fock_+ \cong \Fock_{[0,t[} \ot \Fock_{[t,\infty[}$ via $\e{f}
\longleftrightarrow \e{f|_{[0,t[}} \ot \e{f|_{[t,\infty[}}$. All
cocycles in this paper will be assumed to satisfy~\eqref{adapted}.

Continuity was not given as part of the definition; the next result
mirrors/relies on the corresponding result in semigroup theory.

\begin{propn} \label{coc cts}
Let $X = (X_t)_{t \ges 0}$ be a left cocycle on $\init \ot
\Fock_+$. If $X_t \Tends I$ weakly as $t \Tends 0$ then there are
constants $M, a \in \Real$ such that
\begin{equation}
\label{exp growth}
\norm{X_t} \les Me^{at} \quad \text{for all } t \ges 0.
\end{equation}
Moreover, the map $t \mapsto X_t$ is strongly continuous.
\end{propn}

Cocycles satisfying these continuity conditions will be called
\dfn{$C_0$-cocycles}.

\begin{proof}
Weak convergence to $I$ implies, via two applications of the
Banach-Steinhaus Theorem, that $t \mapsto \norm{X_t}$ is bounded in a
neighbourhood of $0$. The existence of $M$ and $a$ then follows by a
standard argument (see, for example, Proposition~1.18 of~\cite{ebDc0})
since $\norm{X_{r+t}} \les \norm{X_r} \norm{X_t}$, because each
$\si_t$ is contractive.

Weak continuity at $0$, and the local uniform bound for $X$
from~\eqref{exp growth} imply that $X_t \ot I_- \Tends I_{\init \ot
\Fock_\Real}$ weakly, where $I_-$ is the identity on $\Fock_- :=
\Fock_{]{-\infty},0[}$, and we use $\Fock_\Real \cong \Fock_+ \ot
\Fock_-$. If $(\ol{S}_t)_{t \ges 0}$ denotes the strongly continuous
family of \ti{unitary} right shifts on $\init \ot \Fock_\Real$ defined
analogously to the isometries $S_t$, then $Y_t := (X_t \ot I_-)
\ol{S}_t$ is weakly convergent to $I_{\init \ot \Fock_\Real}$.
Moreover, for any $Z \in B(\init \ot \Fock_+)$
\[
\si_t (Z) \ot I_- = \ol{S}_t (Z \ot I_-) \ol{S}^*_t,
\]
and it readily follows that $(Y_t)_{t \ges 0}$ is a semigroup on
$\init \ot \Fock_\Real$. Hence it is strongly continuous, by
Proposition~1.23 of~\cite{ebDc0}, thus so is $t \mapsto X_t \ot I_- =
Y_t \ol{S}^*_t$, and the result follows.
\end{proof}

\begin{rems}
(i) A result in the same spirit is Proposition~2.5
of~\cite{wActsFockI} (reappearing as Proposition~2.3.1
of~\cite{wAEbook}).  It is more general on the one hand since it only
assumes measurability of the cocycle, which is defined with respect to
a \ti{general} $E_0$-semigroup on a von Neumann algebra. However
there are separability assumptions, and essential use is made of the
more restrictive hypothesis that the cocycle be isometric. Similarly,
assumed contractivity of the cocycle is a necessary ingredient of the
alternative proof of the above result for Fock-adapted cocycles given
in Lemma~1.2 of~\cite{FFviaAK}.

(ii) The result extends immediately to right cocycles by use of the
time-reversal operators $\rho_t$ --- see Lemma~\ref{left to right}
below.
\end{rems}

Let $\step = \Lin \{ d \ind_{[0,t[}: d \in \noise, t \ges 0\}$, the
subspace of $L^2(\Rplus; \noise)$ consisting of right continuous, 
piecewise constant functions. It is a dense subspace, so $\Exp := 
\Lin \{\e{f}: f \in \step\}$ is dense in $\Fock_+$. Consequently 
bounded operators on $\init \ot \Fock_+$ are determined by their 
inner products against vectors of the form $u\e{f}$ for $u \in 
\init$, $f \in \step$.

The next result (essentially Proposition~6.2 of~\cite{father}) follows
immediately from adaptedness and~\eqref{sig defn}.

\begin{thm} \label{coc char}
Let $X$ be a bounded adapted process. The following are
equivalent\tu{:}
\begin{rlist}
\item
$X$ is a left cocycle.
\item
For each pair $c,d \in \noise$, $(Q^{c,d}_t := E^{\e{c\ind_{[0,t[}}}
X_t E_{\e{d\ind_{[0,t[}}})_{t \ges 0}$ is a semigroup on $\init$, and
for all $f,g \in \step$
\begin{equation} \label{semi decomp}
E^{\e{f\ind_{[0,t[}}} X_t E_{\e{g\ind_{[0,t[}}} = Q^{f(t_0),
g(t_0)}_{t_1 -t_0} \cdots Q^{f(t_n), g(t_n)}_{t-t_n}
\end{equation}
where $\{0 = t_0 \les t_1 \les \cdots \les t_n \les t \}$ contains the
discontinuities of $f\ind_{[0,t[}$ and $g\ind_{[0,t[}$.
\end{rlist}
If in~\tu{(i)} we replace left by right then~\eqref{semi decomp} 
in~\tu{(ii)} must be replaced by
\begin{equation*}  \label{semi decomp right}
E^{\e{f\ind_{[0,t[}}} X_t E_{\e{g\ind_{[0,t[}}} = Q^{f(t_n),
g(t_n)}_{t-t_n} \cdots Q^{f(t_0), g(t_0)}_{t_1 -t_0}. \tag*{(\ref{semi
decomp})$'$}
\end{equation*}
\end{thm}

The collection of semigroups $\{\Qcd: c,d \in \noise\}$ is the family
of \dfn{associated semigroups} of the cocycle $X$. Since the map
$(c,d) \mapsto \Qcd_t$ is jointly continuous, a cocycle $X$ is
determined by the operators $\Qcd_t$ for $c$ and $d$ taken from a
dense subset of $\noise$. In fact this observation can be further
refined by using totality results such as those contained
in~\cites{PStotal,mStotal,jmLGreif} to show that it is sufficient to
take $c$ and $d$ from a \ti{total} subset of $\noise$ that contains
$0$. If $X$ is a $C_0$-cocycle then it is clear that all of the
associated semigroups are strongly continuous, since the map $x
\mapsto E_x$ is isometric. Conversely if all (or, rather, sufficiently
many) of the associated semigroups are strongly continuous and if $X$
is locally uniformly bounded, then from~\eqref{semi decomp} it follows
that $t \mapsto X_t$ is weakly continuous at $0$, and hence $X$ is a
$C_0$-cocycle. The a priori assumption of local boundedness is
needed here for this (perhaps naive) method of proof to get weak
continuity on all of the complete space $\init \ot \Fock_+$, rather
than just $\init \uot \Exp$, then Proposition~\ref{coc cts} can be
invoked to obtain the improved bound~\eqref{exp growth}.

A stronger hypothesis on the map $t \mapsto X_t$ is
\dfn{Markov-regularity}, as considered in~\cite{father}, which insists
on norm continuity of the \dfn{Markov semigroup} $Q^{0,0}$. For a
$C_0$-cocycle (or, indeed, any locally uniformly bounded cocycle) this
is equivalent to assuming that \ti{all} of the associated semigroups
are norm continuous, which follows easily from the estimate
$\norm{\e{a\ind_{[0,t[}} - \e{c\ind_{[0,t[}}} = O(t^{1/2})$.

In many cases, for instance when proving Theorem~\ref{conisocoi}, it
can be useful to pass from left to right cocycles or vice versa. Two
methods for doing this are taking adjoints and time-reversal, where
for any process $X$ we define $\wt{X}$ by
\[
\wt{X}_t := \rho_t (X_t)
\]
with $\rho_t$ is defined in~\eqref{rho defn}.

\begin{lemma} \label{left to right}
Let $X$ be a bounded adapted process. The following are
equivalent\tu{:}
\begin{rlist}
\item
$X$ is a left cocycle.
\item
$X^* := (X^*_t)_{t \ges 0}$ is a right cocycle.
\item
$\wt{X}$ is a right cocycle.
\end{rlist}
\end{lemma}

\begin{proof}
Equivalence of~(i) and~(ii) is immediate since $\si_t$ is 
${}^*$-homomorphic. Equivalence of~(i) and~(iii) follows from 
Theorem~\ref{coc char} and the fact that $R_t \e{c \ind_{[0,t[}} = 
\e{c \ind_{[0,t[}}$, so that $X$ and $\wt{X}$ share the same family 
of associated semigroups.
\end{proof}

Given a cocycle $X$ we define two unital subalgebras of $B(\init)$:
\begin{align}
\Alg_X &= \text{the norm-closed algebra generated by } \{\Qcd_t: c,d
\in \noise, t \ges 0\},  \label{cpt alg} \\
\intertext{and}
\vM_X &= \text{the von Neumann algebra generated by } \Alg_X.
\label{cpt vNA}
\end{align}
These algebras enter into characterisations of various properties of $X$.
In the language of~\cite{CBonOS} it follows from~\eqref{semi decomp} 
(or~\ref{semi decomp right}) that $X_t \in \Mat (\Fock_+; \Alg_X)_\bd 
\subset \Mat (\Fock_+; \vM_X)_\bd = \vM_X \ot B(\Fock_+)$, where 
$\Mat (\Fock_+; \Vop)_\bd$ denotes the $\Fock_+$-matrix space over an 
operator space $\Vop$.

\begin{propn} \label{coc transfs}
Let $X$ be a left cocycle. We have the following sets of
equivalences\tu{:}
\begin{alist}
\item
\begin{rlist}
\item
$X$ is also a right cocycle.
\item
$X = \wt{X}$.
\item
$\Alg_X$ is commutative.
\end{rlist}
\item
\begin{rlist}
\item
$X^* = \wt{X}$.
\item
$(\Qcd_t)^* = Q^{d,c}_t$ for all $c,d \in \noise$ and $t \ges 0$.
\end{rlist}
\noindent
In this case the algebra $\Alg_X$ is closed under taking adjoints.
\item
\begin{rlist}
\item
$X$ is a self-adjoint cocycle.
\item
$(\Qcd_t)^* = Q^{d,c}_t$ for all $c,d \in \noise$ and $t \ges 0$, and
$\vM_X$ is commutative.
\end{rlist}
\noindent
In this case $X = \wt{X}$ as well.
\end{alist}
\end{propn}

\begin{proof}
(a) This is immediate from Theorem~\ref{coc char} and Lemma~\ref{left
to right} since $X$ is also a right cocycle if and only if not
only~\eqref{semi decomp} but also~\ref{semi decomp right} holds.

\medskip\noindent
(b) $X$ is adapted and $\{\e{f}: f \in \step\}$ is total in $\Fock_+$, 
so $X^* = \wt{X}$ if and only if
\[
\bigl( E^{\e{g\ind_{[0,t[}}} X_t E_{\e{f\ind_{[0,t[}}} \bigr)^* =
E^{\e{f\ind_{[0,t[}}} X_t^* E_{\e{g\ind_{[0,t[}}} =
E^{\e{f\ind_{[0,t[}}} \wt{X}_t E_{\e{g\ind_{[0,t[}}}
\]
for all $f,g \in \step$ and $t \ges 0$. The result thus follows from
Theorem~\ref{coc char} and Lemma~\ref{left to right}, since $R_t \e{c
\ind_{[0,t[}} = \e{c \ind_{[0,t[}}$.

\medskip\noindent
(c\,i \Implies c\,ii) If $X$ is self-adjoint then it is also a right
cocycle by Lemma~\ref{left to right}, thus $\Alg_X$ is commutative
by~(a\,iii), and so $X = X^* = \wt{X}$ by~(a\,ii), so part~(b)
applies, which in particular shows that $\vM_X$ is commutative.

\medskip\noindent
(c\,ii \Implies c\,i) Commutativity of $\vM_X$ implies commutativity
of $\Alg_X$, hence from~(a) we have $X = \wt{X}$, and by~(b) we have
$X^* = \wt{X}$.
\end{proof}

\begin{rems}
Commutativity of $\Alg_X$ does not imply commutativity of $\vM_X$. To
see this take any $A \in B(\init)$ then $(X_t = e^{tA} \ot 
I_{\Fock_+})_{t \ges 0}$ is both a left and right cocycle, since 
$\si_r (X_t) = X_t$ for all $r, t \ges 0$. Moreoever $\Qcd_t = e^{t 
(A+\ip{c}{d})}$, so that $\Alg_X$ is the unital algebra generated by 
$A$, which is certainly commutative, whereas $\vM_X$ is the von 
Neumann algebra generated by $A$, so commutative if and only if $A$ is 
normal.

This result also illustrates some relations that exist between the
algebras defined through~\eqref{cpt alg} and~\eqref{cpt vNA}: for any
cocycle $X$ we have
\[
\Alg_X = \Alg_{\smash{\wt{X}}}, \ \text{ and } \ \vM_X = 
\vM_{\smash{\wt{X}}} = \vM_{X^*},
\]
but $\Alg_X$ need not equal $\Alg_{X^*}$. Other remarks on the
differences between parts~(a),~(b) and~(c) of the proposition are best
made with reference to the stochastic generator of the cocycle, the
subject of the next section, and so are postponed until then.
\end{rems}

\section{Generated cocycles}

A major source of operator cocycles on Fock space comes from solutions
of the left and right Hudson-Parthasarathy QSDEs:
\begin{align*}
dX_t &= X_t F \, d\Lambda_t, \qquad X_0 = I, \tag{L} \\
dX_t &= F X_t \, d\Lambda_t, \qquad X_0 = I. \tag{R}
\end{align*}
Here the coefficient $F$ is a bounded operator on $\init \ot \khat$,
where the use of hats is defined by
\begin{equation}
\label{hats}
\khat := \Comp \op \noise, \qquad \wh{d} = \begin{pmatrix} 1 \\ d
\end{pmatrix} \text{ for } d \in \noise.
\end{equation}
Moreover, let $P_\noise \in B(\khat)$ denote the projection $\khat 
\To \noise$, and $\Delta := I_\init \ot P_\noise$. Since $\init \ot 
\khat \cong \init \op (\init \ot \noise)$, any $F \in B(\init \ot 
\khat)$ can and will be written as
\[
F = \begin{bmatrix} A & B \\ C & D -I_{\init \ot \noise} \end{bmatrix}
\]
for $A \in B(\init)$, $B, C^* \in B(\init \ot \noise; \init)$ and $D
\in B(\init \ot \noise)$.

Straightforward Picard iteration arguments (\cite{jmLGreif}) produce
solutions $\XF = (\XF_t)_{t \ges 0}$ of the left equation~(L) and $\FX
= (\FX_t)_{t \ges 0}$ of the right equation~(R), although neither need
be composed of bounded operators. However the solutions have domain
$\init \uot \Exp$ and satisfy a property called \dfn{weak regularity},
a property shared by any locally bounded process. Moreover $\XF$ and
$\FX$ are the unique weakly regular (weak) solutions to~(L) and~(R) 
for the given $F$. On the other hand, any weakly regular process $X$
satisfies~(L) (or~(R)) (weakly) for at most one $F$. See Theorems~3.1 
and~7.13 of~\cite{mother}. It follows that $(\XF)^*$ is the unique 
weakly regular solution of~(R) for $F^*$, i.e.\ $(\XF)^*|_{\init \uot 
\Exp} = {}^{F^*} \! X$.

The solution $\XF$ enjoys a semigroup decompositions of the
form~\eqref{semi decomp}, where now
\begin{equation} \label{semi gen}
\Qcd_t = e^{tZ^c_d} \ \text{ for } \ Z^c_d = E^{\wh{c}} (F +\Delta)
E_{\wh{d}} = E^{\wh{c}} F E_{\wh{d}} +\ip{c}{d}.
\end{equation}
The solution $\FX$ of~(R) has a similar description involving the same
semigroups, but with the product as in~\ref{semi decomp right}. One
consequence is that a weakly regular process $X$ solves~(L) if and
only if the time-reversed process~$\wt{X}$ satisfies~(R). More
importantly, \ti{if} the solution to~(L) (respectively to~(R)) is a
bounded process then it is a Markov-regular left (resp.\ right)
cocycle. However it is still an open problem to determine all the
operators $F$ that yield bounded solutions. For contractive,
isometric, coisometric, and hence unitary solutions the situation is
understood much better, with the answer being given in terms of the
map $\chi$ on $B(\init \ot \khat)$ where
\begin{equation} \label{cont con}
\chi(F) := F +F^* +F^* \Delta F.
\end{equation}

\begin{thm}[\cites{fFqp8,mother}] \label{conisocoi}
Let $F \in B(\init \ot \khat)$. We have the following sets of
equivalences\tu{:}
\begin{align*}
&\ttu{(a)}
&&\ttu{\hspace*{-3mm}(i) } \FX \text{ is contractive}
&&\ttu{\hspace*{-2mm}(ii) } \XF \text{ is contractive}
&&\ttu{\hspace*{-2mm}(iii) } \chi(F) \les 0 
&&\ttu{\hspace*{-2mm}(iv) } \chi(F^*) \les 0 && \\
&\ttu{(b)}
&&\ttu{\hspace*{-3mm}(i) } \FX \text{ is isometric}
&&\ttu{\hspace*{-2mm}(ii) } \XF \text{ is isometric}
&&\ttu{\hspace*{-2mm}(iii) } \chi(F) = 0 \\
&\ttu{(c)}
&&\ttu{\hspace*{-3mm}(i) } \FX \text{ is coisometric}
&&\ttu{\hspace*{-2mm}(ii) } \XF \text{ is coisometric}
&&\ttu{\hspace*{-2mm}(iii) } \chi(F^*) = 0
\end{align*}
\end{thm}

If, instead, we start with a Markov-regular $C_0$-cocycle $X$ then all
of its associated semigroups $\Qcd$ are norm continuous and so have
bounded generators $Z^c_d$. Let $I$ be a set not containing $0$, set
$\wh{I} = I \cup \{0\}$, and let $\{e_\al\}_{\al \in \wh{I}}$ be an 
orthonormal basis of $\khat$ with $e_0 = \bigl( \begin{smallmatrix} 1 
\\ 0 \end{smallmatrix} \bigr)$, so that $\{e_i\}_{i \in I}$ is an 
orthonormal basis of $0 \op \noise \cong \noise$. This basis induces 
the second of the following isomorphisms:
\begin{equation} \label{direct sum}
\init \ot \khat \cong \init \op (\init \ot \noise) \cong \init \op
\bigoplus\ds{^{(\dim \noise)}} \init.
\end{equation}
Now define operators $\{F^\al_\be: \al, \be \in \wh{I}\} \subset 
B(\init)$ through
\begin{equation} \label{F cpts}
\begin{aligned}
F^0_0 &= Z^{e_0}_{e_0}, \quad F^i_0 = Z^{e_i}_{e_0} - Z^{e_0}_{e_0},
\quad F^0_j = Z^{e_0}_{e_j} - Z^{e_0}_{e_0} \quad \text{ and} \\
F^i_j &= Z^{e_i}_{e_j} - Z^{e_i}_{e_0} - Z^{e_0}_{e_j} + Z^{e_0}_{e_0}
-\delta^i_j I_\init,
\end{aligned}
\end{equation}
for $i,j \in I$, and where $\delta^i_j$ is the Kronecker delta. If
$\noise$ is finite-dimensional, these $F^\al_\be$ can be regarded as
the components of the matrix associated to a bounded operator $F \in
B(\init \ot \khat)$ through~\eqref{direct sum}, and it follows that $X
= \XF$ or $\FX$ as appropriate. That is, the cocycle is the solution
of the relevant QSDE. For infinite-dimensional $\noise$, a priori the 
matrix $[F^\al_\be]$ only gives us a form on $\init \ot \khat$, with
respect to which $X$ satisfies a weak form of~(L) or~(R) --- this is
Theorem~6.6 of~\cite{father}. However, if the cocycle is in addition
\ti{contractive} then the form is bounded, and so the $F^\al_\be$ are
the components of some $F \in B(\init \ot \khat)$ as before.

Recall the subalgebras $\Alg_X$ and $\vM_X$ of $B(\init)$ associated
to a cocycle $X$ by~\eqref{cpt alg} and~\eqref{cpt vNA}. For a
generated cocycle $X$, i.e.\ one satisfying~(L) or~(R) for some $F \in
B(\init \ot \khat)$, it follows from~\eqref{semi gen} and~\eqref{F 
cpts} that $\Alg_X$ is the unital algebra generated by the components
$F^\al_\be$ of $F$. That is, $F \in \Mat (\khat; \Alg_X)_\bd$, the
$\khat$-matrix space over $\Alg_X$. Moreover, from~\eqref{semi gen}
and~\eqref{F cpts} we have
\[
(\Qcd_t)^* = Q^{d,c}_t \ \text{ for all } c,d \in \noise, t \ges 0
\ \Iff \ F = F^*.
\]
The following is thus the infinitesimal version of
Proposition~\ref{coc transfs}:

\begin{propn} \label{tisa gen}
Let $F \in B(\init \ot \khat)$ and suppose that $\XF$ is bounded with
locally uniform bounds, hence a Markov-regular left $C_0$-cocycle. We
have the following sets of equivalences\tu{:}
\begin{alist}
\item
\begin{alist}
\item
$\XF = \wt{\XF} = \FX$.
\item
$F \in \Mat(\khat; \Clg)_\bd$ for some commutative subalgebra $\Clg
\subset B(\init)$.
\end{alist}
\item
\begin{alist}
\item
$\wt{\XF} = (\XF)^*$.
\item
$F = F^*$.
\end{alist}
\noindent
In this case $\Alg_X$ is closed under taking adjoints.
\item
\begin{alist}
\item
$\XF$ is self-adjoint.
\item
$F = F^*$ and $F \in \vN \ot B(\khat)$ for some commutative von
Neumann algebra $\vN$.
\end{alist}
\end{alist}
\end{propn}

\begin{egs}
(i) If $\init = \noise = \Comp$ and $F = \begin{sbmatrix} -1/2 & -1
\\ 1 & 0 \end{sbmatrix}$ then $\Alg_X = \vM_X = \Comp$, so $\XF = 
\wt{\XF} = \FX$, i.e.\ $\XF$ is both a left and right cocycle. 
However $F \neq F^*$, and so~(a) neither implies~(b) nor~(c). 
Furthermore, $\Alg_X$ being closed under taking adjoints does not 
imply $F = F^*$. In this example $\XF_t = W(\ind_{[0,t[})$, the Weyl 
operator associated to $\ind_{[0,t[} \in L^2(\Rplus)$.

(ii) As noted after Proposition~\ref{coc transfs}, commutativity of 
$\Alg_X$ does not imply commutativity of $\vM_X$. This also shows 
that~(a) does not imply~(c).

(iii) Let $F = \begin{sbmatrix} A & B \\ C & D-I \end{sbmatrix}$ 
with $A = A^* = -\frac{1}{2} C^*C$, $B = C^*$, $D = D^*$, $D^2 = I$ 
and $(I+D)C = 0$. Then $F = F^*$ and $\chi(F) = 0$, so from 
Theorem~\ref{conisocoi} we have that $\XF$ is unitary, and from~(b) 
of the above proposition that $(\XF)^* = \wt{\XF}$. However if we 
ensure that $\Alg_X$ is not commutative then $(\XF)^* \neq \XF$; 
this can be achieved by taking $\init = \Comp^2$, $\noise = \Comp$ 
and $C = \begin{sbmatrix} 1 & 0 \\ -1 & 0 \end{sbmatrix}$, $D = 
\begin{sbmatrix} 0 & 1 \\ 1 & 0 \end{sbmatrix}$. This shows that~(b) 
implies neither~(a) nor~(c).
\end{egs}

In the examples above it is the algebra generated by the
components of $F$ rather than the von Neumann algebra $\vN_F$
generated by $F$ itself that is of interest. For example in~(i) 
$\vM_X$ is commutative, whereas $\vN_F$ is not, and the opposite holds 
true in the example in~(iii).

\begin{thm} \label{pos gen}
Let $F \in B(\init \ot \khat)$ and suppose that $\XF$ is bounded with
locally uniform bounds, hence a Markov-regular left $C_0$-cocycle. The
following are equivalent\tu{:}
\begin{rlist}
\item
$\XF_t \ges 0$ for all $t \ges 0$.
\item
$F = F^* \in \vN \ot B(\khat)$ for some commutative von Neumann
algebra $\vN$, and $\Delta F \Delta +\Delta \ges 0$.
\end{rlist}
\end{thm}

\begin{proof}
Given any von Neumann algebra $\vN$ and $F \in \vN \ot B(\khat)$, if
we define $\theta: \vN \To \vN \ot B(\khat)$ by $\theta (a) = F(a \ot
I_\khat)$ and assume that $\XF$ is a bounded solution of~(L) then the
mapping process $k_t (a) := \XF_t (a \ot I_{\Fock_+})$ is a solution
of the Evans-Hudson QSDE $dk_t = k_t \circ \theta \, d\Lambda_t$.
Moreover, by Theorem~4.1 of~\cite{mother}, $k$ is completely positive
if and only if
\begin{equation} \label{CP gen}
\theta (a) = \psi (a) +E_{\wh{0}}\, a K + K^* a E^{\wh{0}} - a \ot
P_\noise
\end{equation}
for some completely positive map $\psi: \vN \To \vN \ot B(\khat)$ and
$K \in B(\init \ot \khat; \init)$.

\medskip
\noindent
(i \Implies\ ii) Suppose that each $\XF_t$ is positive, then $F \in
\vN \ot B(\khat)$ for some commutative von Neumann algebra $\vN$ by
Proposition~\ref{tisa gen}, hence $(\XF_t)^{1/2} \in \vN \ot
B(\Fock_+)$, and so commutes with $a \ot I_{\Fock_+}$, showing that
the flow $k$ is completely positive. In particular, since $\Delta
E_{\wh{0}} = 0$,
\[
\Delta F \Delta +\Delta = \Delta \theta (1) \Delta +\Delta = \Delta
\psi (1) \Delta \ges 0.
\]

\medskip
\noindent
(ii \Implies\ i) Write $F = \begin{sbmatrix} A & B \\ B^* & D-I
\end{sbmatrix}$ so that $D \ges 0$. Since $\vN$ is commutative,
\[
\psi (a) := \begin{bmatrix} 0 & 0 \\ 0 & D (a \ot I_\noise)
\end{bmatrix} = \begin{bmatrix} 0 & 0 \\ 0 & D^{1/2} (a \ot I_\noise)
D^{1/2} \end{bmatrix}
\]
is completely positive. Setting $K = \begin{bmatrix} \frac{1}{2} A & B 
\end{bmatrix}$ we get
\[
\psi (a) +E_{\wh{0}}\, a K +K^* a E^{\wh{0}} -a \ot P_\noise = F (a
\ot I_\khat),
\]
so $\theta$ has the form~\eqref{CP gen}, and thus generates a
completely positive flow. In particular $\XF_t = k_t (1)$ must be
positive.
\end{proof}

\begin{cor} \label{cpos gen}
Let $F \in B(\init \ot \khat)$. The following are equivalent\tu{:}
\begin{rlist}
\item
$\XF$ is a positive contraction cocycle.
\item
$F \in \vN \ot B(\khat)$ for some commutative von Neumann algebra
$\vN$, $F \les 0$ and $\Delta F \Delta +\Delta \ges 0$.
\item
$F = \begin{sbmatrix} A & B \\ B^* & D-I \end{sbmatrix} \in \vN \ot
B(\khat)$ for some commutative von Neumann algebra $\vN$, with $A \les
0$, $0 \les D \les I$ and $B = (-A)^{1/2} V (I-D)^{1/2}$ for some
contraction $V \in B(\init \ot \noise; \init)$.
\end{rlist}
\end{cor}

\begin{proof}
If the flow $k$ generated by $\theta(a) = F(a \ot I_\khat)$ is
completely positive then it is contractive if and only if $\theta (1)
= F \les 0$ (\cite{LPgran}*{Theorem~5.1} or
\cite{mother}*{Proposition~5.1}). Moreover if $k$ is positive then
$\norm{k_t} = \norm{k_t (1)} = \norm{X_t}$. This gives the equivalence
of~(i) and~(ii). Part~(iii) follows from a standard characterisation
of positive $2 \times 2$ operator matrices (e.g.\
~\cite{dilation}*{Lemma~2.1}).
\end{proof}

\begin{rem}
In terms of the operators in~(iii) one may recognise Bhat's
characterisation of positive contraction cocycles in the special case 
when $\init = \Comp$ (\cite{bvrBCCR}*{Theorem~7.5}). His focus there
was on \dfn{local cocycles}, that is cocycles which satisfy $X_t \in 
\si_t \bigl( B(\init \ot \Fock_+) \bigr)'$ for all $t \ges 0$. For the
CCR flow
\[
\si_t \bigl( B(\init \ot \Fock_+) \bigr) = B(\init) \ot
I_{\Fock_{[0,t[}} \ot B(\Fock_{[t,\infty[}),
\]
so this assumption is stronger than mere adaptedness, and forces $X_t$ 
to act trivially on $\init$, equivalently we must have $\Alg_X = 
\Comp$, or $F \in I_\init \ot B(\khat)$. Hence one may restrict to the 
case $\init = \Comp$ without loss of generality.
\end{rem}

The final characterisations rely on being able to multiply cocycles
together to produce new cocycles.

\begin{lemma} \label{products}
Let $F,G \in B(\init \ot \khat)$ and suppose that the solutions $\XF$
and $X^G$ to~\tu{(L)} for these coefficients are both bounded with
locally uniform bounds. Assume also that
\begin{equation} \label{commute}
(F \ot I_{\Fock_+}) \wh{X^G_t} = \wh{X^G_t} (F \ot I_{\Fock_+}) \quad
\text{ for all } t \ges 0
\end{equation}
where $\wh{X^G_t} \in B(\init \ot \khat \ot \Fock_+)$ denotes the
result of ampliating $X^G_t$ to $\init \ot \khat \ot \Fock_+$. In this 
case the product $\XF X^G$ is a bounded left $C_0$-cocycle with 
stochastic generator $F +G +F \Delta G$.
\end{lemma}

\begin{proof}
The adjoint process $\bigl( (\XF_t)^* \bigr)_{t \ges 0}$ is a right
cocycle with stochastic generator $F^*$, and so the quantum It\^{o}
formula gives
\begin{multline*}
\ip{u\e{f}}{(\XF_t X^G_t -I) v\e{g}} = \\
\begin{aligned}
\int^t_0 \Bigl\{ & \ip{\wh{\XF_s}^* u\wh{f}(s)
\e{f}}{\wh{X^G_s} (G \ot I_{\Fock_+}) v \wh{g}(s) \e{g}} \\
& + \ip{(F^* \ot I_{\Fock_+}) \wh{\XF_s}^* u\wh{f}(s)
\e{f}}{\wh{X^G_s} v \wh{g}(s) \e{g}} \\
& + \ip{(F^* \ot I_{\Fock_+}) \wh{\XF_s}^* u\wh{f}(s) \e{f}}{(\Delta \ot
I_{\Fock_+}) \wh{X^G_s} (G \ot I_{\Fock_+}) v \wh{g}(s) \e{g}} \Bigr\}
\, ds.
\end{aligned}
\end{multline*}
The commutativity assumed in~\eqref{commute} shows that the weakly
regular process $\XF X^G$ satisfies~(L) for $F +G +F \Delta G$, and so
is a cocycle with this generator.
\end{proof}

\begin{rem}
If $F \in \vM \ot B(\khat)$ and $G \in \vN \ot B(\khat)$ for von
Neumann algebras $\vM$ and $\vN$ then a sufficient condition
for~\eqref{commute} is $\vN \subset \vM'$, since $\wh{X^G_t} \in \vN
\ot I_\khat \ot B(\Fock_+)$. In particular this is true if $\vM$ is
commutative and $\vN = \vM$.
\end{rem}

\begin{propn} \label{proj gen}
Let $F \in B(\init \ot \khat)$. The following are equivalent\tu{:}
\begin{rlist}
\item
$\XF$ is an orthogonal projection-valued cocycle.
\item
$F \in \vN \ot B(\khat)$ for some commutative von Neumann algebra
$\vN$, and $F +F^* \Delta F = 0$.
\item
$F = \begin{sbmatrix} -BB^* & B \\ B^* & P-I \end{sbmatrix} \in \vN
\ot B(\khat)$ for some commutative von Neumann algebra $\vN$ where $P
\in \vN \ot B(\noise)$ is an orthogonal projection, and $BP=0$.
\end{rlist}
\end{propn}

\begin{proof}
(i \Implies\ ii) Since $\XF$ is self-adjoint, $F \in \vN \ot
B(\khat)$ for a commutative von Neumann algebra $\vN$, and $F = F^*$. 
It follows that $(F \ot I_{\Fock_+}) \wh{\XF_t} = \wh{\XF_t} (F \ot 
I_{\Fock_+})$, hence $(X^F)^2$ is a cocycle with stochastic generator 
$2F +F \Delta F$ by Lemma~\ref{products}. But we assumed that $\XF_t 
= (\XF_t)^2$, and since generators are unique we get
\[
F = 2F +F \Delta F = 2F +F^* \Delta F
\]
as required.

\medskip
\noindent
(ii \Implies\ i) From $F +F^* \Delta F = 0$ it follows that $F$ is
self-adjoint, and that $F \les 0$. Thus part~(a) of
Theorem~\ref{conisocoi} and part~(c) of Proposition~\ref{tisa gen}
apply to show that $\XF$ is a self-adjoint contraction $C_0$-cocycle.
But now Lemma~\ref{products} applies to show that $(\XF)^2$ is also a
cocycle, with generator $2F +F \Delta F = F$, and so by uniqueness of
solutions to~(L) we have that each $\XF_t$ is an orthogonal
projection.

\medskip
\noindent
(ii \Iff\ iii) Simple algebra.
\end{proof}

The final characterisation rests on equivalences between operator
(in-)equalities involving $\chi(F)$ defined in~\eqref{cont con} and
the following additional functions of $F$:
\begin{align}
\pi(F) &:= F +F^* +F^* \Delta F +F \Delta F +F \Delta F^* +F \Delta
F^* \Delta F \label{pi con} \\
\varphi(F) &:= \chi(F) +\chi(F) \Delta \chi(F). \notag
\end{align}

\begin{lemma} \label{ineqs}
For any $F \in B(\init \ot \khat)$ we have the following sets of
equivalences\tu{:}
\begin{align*}
&\ttu{(a)}
&&\ttu{\hspace*{-4mm}(i) } \chi(F) \les 0
&&\ttu{(ii) } \chi(F^*) \les 0
&&\ttu{(iii) } \varphi(F) \les 0
&&\ttu{(iv) } \varphi(F^*) \les 0 &&\\
&\ttu{(b)}
&&\ttu{\hspace*{-4mm}(i) } \pi(F) = 0
&&\ttu{(ii) } \pi(F^*) = 0
&&\ttu{(iii) } \varphi(F) = 0
&&\ttu{(iv) } \varphi(F^*) = 0 &&
\end{align*}
\end{lemma}

\begin{proof}
(a) Since $\chi(F) = \chi(F)^*$, if $\varphi(F) \les 0$ then $\chi(F)
\les -\chi(F) \Delta \chi(F) \les 0$. Thus (iii) \Implies\ (i) and
(iv) \Implies\ (ii). However note that
\begin{equation} \label{phi id}
\varphi(F) = (I +F^* \Delta) \chi(F^*) (I +\Delta F),
\end{equation}
from which it follows that (ii) \Implies\ (iii) and (i) \Implies\ 
(iv).

\medskip
\noindent
(b) Now $\pi(F^*) = \pi(F)^*$ so (i) \Iff\ (ii). Also $\pi(F) =
\chi(F^*) (I +\Delta F)$, hence (i) \Implies\ (iii) by~\eqref{phi id}.
Finally, if $\varphi(F) =0$ then $\chi(F^*) \les 0$ by part~(a), and
so
\[
0 = -\varphi(F) = \bigl[ \bigl( -\chi(F^*) \bigr)^{1/2} (I +\Delta F)
\bigr]^* \bigl[ \bigl( -\chi(F^*) \bigr)^{1/2} (I +\Delta F) \bigr]
\]
giving (iii) \Implies\ (i).
\end{proof}

\begin{propn} \label{pi gen}
Let $F \in \vN \ot B(\khat)$ for a commutative von Neumann algebra
$\vN$. The following are equivalent\tu{:}
\begin{rlist}
\item
$\XF$ is a partial isometry-valued cocycle.
\item
$\pi(F) = 0$, where $\pi(F)$ is defined in~\eqref{pi con} above.
\end{rlist}
\end{propn}

\begin{proof}
If $\XF$ is partial isometry-valued then since $\vN$ is commutative,
the cocycle $(\XF)^* \XF$ is projection-valued with generator $\chi(F)
= F +F^* +F^* \Delta F$ by Lemma~\ref{products}.  Hence $\varphi(F) =
\chi(F) +\chi(F) \Delta \chi(F) = 0$ by Proposition~\ref{proj gen} and
so $\pi(F) = 0$ by the lemma above.

Conversely, if $\pi(F) = 0$ then $\varphi(F) = 0$ by the lemma, hence
$\chi(F) \les 0$ and so $\XF$ is a contraction cocycle by part~(a) of
Theorem~\ref{conisocoi}. Again Lemma~\ref{products} can be invoked to
show that $(\XF)^* \XF$ is a (bounded) cocycle with generator
$\chi(F)$ which satisfies the conditions of Proposition~\ref{proj gen}
and hence is projection-valued, so that $\XF$ is itself a partial
isometry-valued cocycle.
\end{proof}

The condition $\pi(F)=0$ is necessarily satisfied by the generator of
\ti{any} Markov-regular partial isometry-valued cocycle, as can be
shown by standard independence of quantum stochastic
integrators/differentiation at zero arguments. In particular if $F =
\begin{sbmatrix} 0 & 0 \\ 0 & D-I \end{sbmatrix}$ then $\pi(F) =0$ if 
and only if $D$ is a partial isometry, but for such pure-gauge 
cocycles this condition is in general \ti{not} sufficient to imply 
that $\XF$ is partial isometry-valued, as can be seen by using the 
explicit solution of~(L) given in~\cite{jmLGreif}*{Example~5.3}. For 
each  $n \in \Nat$ and $1 \les j \les n$ define $D^{(n)}_j \in B(\init 
\ot \noise^{\ot n})$ by having $D$ act on $\init$ and the $j$th copy 
of $\noise$, and ampliating to the other copies of $\noise$. Set
\[
D^{(n)} := D^{(n)}_1 \cdots D^{(n)}_n \in B(\init \ot \noise^{\ot n}).
\]

\begin{propn} \label{not suff}
Let $D \in B(\init \ot \noise)$ be a contraction and set $F =
\begin{sbmatrix} 0 & 0 \\ 0 & D-I \end{sbmatrix}$. The following are
equivalent\tu{:}
\begin{rlist}
\item
$\XF$ is a partial isometry-valued cocycle.
\item
$D^{(n)}$ is a partial isometry for each $n \in \Nat$.
\end{rlist}
\end{propn}

\begin{proof}
The symmetric tensor product of $n$ copies of $L^2 ([0,t[; \noise)$
can be naturally identified with $L^2 (\Delta^n_t; \noise^{\ot n})$,
where $\Delta^n_t = \{0 < t_1 < \cdots < t_n < t\} \subset (\Rplus)^n$
(see~\cite{mSwhitebialg} or~\cite{jmLGreif} for details). It follows
that
\[
\init \ot \Fock_+ \cong \biggl( \bigoplus_{n=0}^\infty L^2 (\Delta^n_t;
\init \ot \noise^{\ot n}) \biggr) \ot \Fock_{[t,\infty[}
\]
and that under this identification the solution $\XF$ of~(L) has the
explicit form
\[
\XF_t = \biggl( \bigoplus_{n=0}^\infty I_{L^2 (\Delta^n_t)} \ot
D^{(n)} \biggr) \ot I_{\Fock_{[t,\infty[}},
\]
(see~\cite{jmLGreif}). The result follows.
\end{proof}

\begin{eg}
As a special case, if $\noise = \Comp$ then $\init \ot \noise^{\ot n}
\cong \init$ and $D^{(n)} = D^n$, the usual $n$th power of $D$. Thus
in this setting $\XF$ is partial isometry-valued if and only if $D^n$
is a partial isometry for each $n$, whereas $\pi (F) =0$ merely if $D$
alone is a partial isometry. If we take $\init = \Comp^2$ and the
partial isometry
\[
D = 
\begin{bmatrix}
\cos \theta & 0 \\ \sin \theta & 0
\end{bmatrix}, \qquad \theta \notin \pi \Int/2,
\]
then $D^2 (D^2)^* D^2 \neq D^2$, so that $\XF$ is \ti{not} partial
isometry-valued.

An operator $D$ such that $D^n$ is a partial isometry for each $n \in 
\Nat$ is called \dfn{power partial isometry}; these have been
characterised by Halmos and Wallen (\cite{HWpowerPIs}).
\end{eg}

\section{A transformation of positive cocycles}

\begin{propn} \label{powers}
Let $X = (X_t)_{t \ges 0}$ be a left cocycle with $X_t \ges 0$ for
each $t \ges 0$. Then for each real number $\al > 0$ the family $X^\al
= (X^\al_t)_{t \ges 0}$ is a left cocycle.
\end{propn}

\begin{proof}
We are dealing with a self-adjoint cocycle, so it is both a left and a
right cocycle, hence
\begin{equation} \label{op comm}
X_{r+t} = X_r \si_r (X_t) = \si_r (X_t) X_r \quad \text{for all } r,t
\ges 0.
\end{equation}
From this it is clear that $X^n$ is a cocycle for any integer $n \ges 
1$, and one consisting of positive operators.

Since $\si_r$ is a ${}^*$-homomorphism it follows that $\si_r (X_t) 
\ges 0$ and that $\si_r (X_t^{1/2}) = \si_r (X_t)^{1/2}$. Moreover
from~\eqref{op comm} and the continuous functional calculus we obtain
\[
[X_r^{1/2}, \si_r (X_t)^{1/2}] = [X^{1/2}_r, \si_r (X^{1/2}_t)] = 0,
\]
and thus
\[
X^{1/2}_r \si_r (X^{1/2}_t) \ges 0, \qquad \bigl( X^{1/2}_r \si_r
(X^{1/2}_t) \bigr)^2 = X_{r+t}.
\]
Hence, by uniqueness of positive square roots, $X^{1/2}$ is a left
and right cocycle of positive operators.

These two observations show that $X^\al$ is a cocycle for any dyadic
rational $\al$. To get the desired result for any $\al > 0$ let
$(\al_n)_{n \ges 1}$ be a decreasing sequence of dyadic rationals with
$\al_n \Tends \al$. Now if $h_\be (t) := t^\be$ for $\be > 0$ then
$h_{\al_n} \Tends h_\al$ locally uniformly --- the function sequence
is pointwise increasing on $[0,1]$ and pointwise decreasing on $[1,T]$
for any $T > 1$, and so Dini's Theorem may be applied. Thus appealing
to the continuous functional calculus once more and continuity of
$\si_r$ is enough to show that $X^\al_{r+t} = X^\al_r \si_r (X^\al_t)$
as required.
\end{proof}

It should be noted that the proof of the above result does not on
particular properties of the CCR flow $\si$ on Fock space. Indeed, the 
result is valid for any $E$-semigroup since even preservation of the 
identity by $\si$ is not used.

However, if the cocycle $X$ is a Markov-regular positive contraction
cocycle then it has a stochastic generator $F$. The next results
discuss how $F$ is transformed by taking powers of $X$, and this is
mediated through the following functions from the algebra $C[0,1]$,
defined for each $\al > 0$.
\[
f_\al(t) = \begin{cases} \dfrac{\al-1 -\al t +t^\al}{(1-t)^2} &
\text{if } t < 1, \\ \frac{1}{2} \al(\al -1) & \text{if } t=1,
\end{cases}
\quad
g_\al(t) = \begin{cases} \dfrac{1 -t^\al}{1-t} & \text{if } t < 1, \\ 
\al & \text{if } t=1, \end{cases}
\quad
h_\al(t) = t^\al.
\]
Note that $h_\al$ is a homeomorphism $[0,1] \To [0,1]$, so induces an
automorphism of $C[0,1]$ by composition. Also we have the following
identities, valid for all $\al, \be > 0$:
\begin{subequations}
\begin{gather}
g_\al = \al -(1- h_1) f_\al; \label{for lemma} \\
f_\al +f_\be +g_\al g_\be = f_{\al+\be}, \quad g_\be +g_\al h_\be =
g_{\al+\be}, \quad h_\al h_\be = h_{\al+\be}; \label{add} \\
\be f_\al +g_\al^2 (f_\be \circ h_\al) = f_{\al\be}, \quad g_\al
(g_\be \circ h_\al) = g_{\al\be}, \quad h_\be \circ h_\al =
h_{\al\be};
\label{comp}
\end{gather}
and the inequalities
\begin{equation}
f_\al(t) \les f_\be(t), \quad g_\al(t) \les g_\be(t), \quad h_\al(t)
\ges h_\be(t), \label{monotone}
\end{equation}
\end{subequations}
valid for all $t \in [0,1]$ and $1 \les \al < \be$.

\begin{lemma} \label{Fal gens}
Suppose $F = \begin{sbmatrix} A & B \\ B^* & D-I \end{sbmatrix} \in
B(\init \ot \khat)$ is the generator of a positive contraction
cocycle. Then so is $F_\al \in B(\init \ot \khat)$ where
\begin{equation} \label{Fal}
F_\al := \begin{bmatrix} \al A +B f_\al(D) B^* & B g_\al(D) \\ g_\al(D)
B^* & h_\al(D) -I \end{bmatrix}.
\end{equation}
\end{lemma}

\begin{rem}
The proof uses the following elementary fact: if $\Kil_1$, $\Kil_2$
and $\Kil_3$ are Hilbert spaces, and $S \in B(\Kil_1; \Kil_2)$, $T \in
B(\Kil_1; \Kil_3)$ such that $S^*S \les T^*T$, then there is a
contraction $W \in B(\Kil_2; \Kil_3)$ such that $S = WT$. This follows
since the inequality allows us to define $W$ by setting $W (T\xi) =
S\xi$ on $\Ran T$ and $W|_{(\Ran T)^\perp} = 0$.
\end{rem}

\begin{proof}
By condition~(iii) of Corollary~\ref{cpos gen}, $F \in \vN \ot
B(\khat)$ for some commutative von Neumann algebra $\vN$, $A \les 0$,
$0 \les D \les I$ and there is a contraction $V$ such that $B =
(-A)^{1/2} V (I-D)^{1/2}$.

Now $F_\al \in \vN \ot B(\khat)$, and $0 \les h_\al(D) \les I$ since
$h_\al([0,1]) = [0,1]$. Also $g_\al \ges 0$, so
\begin{align*}
0 &\les (-A)^{1/2} V g_\al(D) V^* (-A)^{1/2} \\
&= (-A)^{1/2} V (\al -(I-D)^{1/2} f_\al(D) (I-D)^{1/2}) V^* (-A)^{1/2}
\\
&\les -\al A -B f_\al(D) B^*,
\end{align*}
using~\eqref{for lemma}. These inequalities prove the existence of a
contraction $W \in B(\init; \init \ot \noise)$ that satisfies
\[
W (-\al A -B f_\al(D) B^*)^{1/2} = g_\al (D)^{1/2} V^* (-A)^{1/2},
\]
and so
\begin{align*}
B g_\al(D) &= (-A)^{1/2} V g_\al(D)^{1/2} (I-D)^{1/2} g_\al(D)^{1/2}
\\
&= (-\al A -B f_\al(D) B^*)^{1/2} W^* (I -h_\al(D))^{1/2},
\end{align*}
since $(I-D) g_\al(D) = I -h_\al(D)$. Thus $F_\al$ satisfies
condition~(iii) of Corollary~\ref{cpos gen}, showing that it is the
generator of a positive contraction cocycle.
\end{proof}

\begin{thm} \label{powergen}
Let $X$ be a Markov-regular positive contraction cocycle with
stochastic generator $F$. Then for each real $\al > 0$ the cocycle
$X^\al$ is Markov-regular with generator $F_\al$ given by~\eqref{Fal}.
\end{thm}

\begin{proof}
The identities~\eqref{add} lead immediately to
\begin{equation} \label{sum}
F_\al +F_\be +F_\al \Delta F_\be = F_{\al+\be} \quad \text{for all }
\al, \be > 0.
\end{equation}
In particular, noting that $F_1 = F$, it follows from
Lemma~\ref{products} and an induction argument that $X^n$ is
Markov-regular and has generator $F_n$ for each $n \in \Nat$.

Next, Lemma~\ref{Fal gens} shows that $\Fh$ is the generator of some
positive contraction cocycle $Y$. By~\eqref{sum} and
Lemma~\ref{products}, $Y^2$ is also a cocycle with generator $2 \Fh
+\Fh \Delta \Fh = F_1 = F$, and so $Y^2 = X$ by uniqueness of
generators.  Thus the cocycle $X^{1/2}$ has generator $\Fh$.

Now the identities~\eqref{comp} give $(F_\al)_\be = F_{\al\be}$
for all $\al, \be > 0$, so square roots may be taken repeatedly, 
followed by taking arbitrarily large integer powers to show that 
$X^\al$ is Markov-regular with generator $F_\al$ for each dyadic 
rational $\al > 0$.

If we choose any real number $\al > 1$ and let $(\al_n)_{n \ges 1}$ be
a sequence of such rationals with $\al_n \downarrow \al$, then the
function sequences $(f_{\al_n})$, $(g_{\al_n})$ and $(h_{\al_n})$
converge pointwise to $f_\al$, $g_\al$ and $h_\al$. The
inequalities~\eqref{monotone} show that the convergence is also
monotonic, and hence uniform by Dini's Theorem, so that $F_{\al_n}
\Tends F_\al$ in norm. It follows from~\eqref{semi decomp}
and~\eqref{semi gen} that the associated semigroups of the cocycle
$X^\al$ are the norm limits of the semigroups associated to
$X^{\al_n}$, and thus $X^\al$ has stochastic generator $F_\al$.

Finally, for any remaining $0 < \al < 1$ pick $n \in \Nat$ so that
$\be := 2^n \al > 1$, then $X^\be$ has generator $F_\be$, and $X^\al =
(X^\be)^{2^{-n}}$ has generator $(F_\be)_{2^{-n}} = F_\al$.
\end{proof}

\section{Polar decomposition}

One obvious question to ask given the results above is the following:
if $X$ is a contraction cocycle such that $\vM_X$ is commutative then
we can form the positive part process $(|X_t|)_{t \ges 0} = \bigl(
(X^*_t X_t)^{1/2} \bigr)_{t \ges 0}$ which is again a cocycle, so can
we choose partial isometries $U_t$ so that $U_t |X_t| = X_t$ for each 
$t$ \ti{and} so that $(U_t)_{t \ges 0}$ is a cocycle?

What follows answers this question when $X$ is Markov-regular with 
generator $F = \begin{sbmatrix} A & B \\ C & D-I \end{sbmatrix}$. The 
necessary  and sufficient conditions on $F$ for contractivity of $X$ 
is $\chi(F)  \les 0$ (Theorem~\ref{conisocoi}) which translates as:
\begin{equation} \label{con conds}
\begin{gathered}
\norm{D} \les 1, \quad A+A^*+C^*C \les 0, \quad \text{ and} \\
B+C^*D = (-A-A^*-C^*C)^{1/2} V (I-D^*D)^{1/2}
\end{gathered}
\end{equation}
for some contraction $V \in B(\init \ot \noise; \init)$.
Lemma~\ref{products}, Proposition~\ref{powers} and
Theorem~\ref{powergen} combine to show that $(|X_t|)_{t \ges 0}$ is a
Markov-regular cocycle with generator $G = \chi(F)_\frac{1}{2}$, which
equals
\[
\begin{bmatrix}
\frac{1}{2} (A+A^*+C^*C) +(B+C^*D) \fh(|D|^2) (B^*+D^*C) &
(B+C^*D) \gh(|D|^2) \\
\gh(|D|^2) (B^*+D^*C) & |D|-I
\end{bmatrix}
\]

Now suppose that $U$ is a partial isometry-valued cocycle with
generator $E = \begin{sbmatrix} K & L \\ M & N-I \end{sbmatrix} \in
\vM_X \ot B(\khat)$. Then the product process $(U_t |X_t|)_{t \ges 0}$
is a cocycle with generator $E+G+E \Delta G$ (Lemma~\ref{products}).
This must equal $F$ to give $U|X| = X$, and thus $K$, $L$, $M$ and $N$
must be chosen to satisfy
\begin{subequations} \label{gen eqns}
\begin{align}
N|D| &= D, \label{N eqn} \\
M &= C -N \gh(|D|^2) (B^*+D^*C), \label{M eqn} \\
L|D| &= B -(B+C^*D) \gh(|D|^2), \label{L eqn} \\
K &= \tfrac{1}{2} (A-A^*-C^*C) -(B+C^*D) \fh(|D|^2) (B^*+D^*C) \label{K
eqn} \\
& \quad -L\gh(|D|^2) (B^*+D^*C). \notag
\end{align}
\end{subequations}
Note that $N$ and $L$ are fixed on $\Ran |D|$ by these equations, and
once these are chosen the operators $K$ and $M$ are defined
by~\eqref{K eqn} and~\eqref{M eqn} respectively. Also, we want $U$ to
be partial isometry-valued, so need to satisfy $\pi(E) = 0$ by
Proposition~\ref{pi gen}. This is equivalent to requiring
\begin{equation} \label{pi conds}
N = NN^*N, \quad M^*N +LN^*N = 0 \ \text{ and } \ K+K^*+M^*M
+L(I-N^*N)L^* = 0
\end{equation}
So now if we choose any partial isometry $N$ that satisfies~\eqref{N
eqn} then one solution to this problem is obtained by setting
\[
L = -C^*N +(B+C^*D) \gh(|D|^2).
\]
Since $\sqrt{t}\, \gh(t) = 1- \gh(t)$ it is easy to check~\eqref{L
eqn}; moreover the second equation in~\eqref{pi conds} follows.
Checking that the third equation holds is much more tedious, but is
greatly assisted by noting that $2\fh +(\gh)^2 = 0$. Summarising the
above we have:

\begin{thm} \label{polar decomp}
Every Markov-regular contraction cocycle $X$ for which $\vM_X$ is
commutative can be written as the product of a partial isometry-valued
cocycle and a positive contraction cocycle.
\end{thm}

\begin{egs}
(i) Take $\init = \Comp$, then $\Alg_X = \vM_X = \Comp$, thus
Theorem~\ref{polar decomp} is applicable to \ti{any} contraction
cocycle on $\Fock_+$, since also every cocycle is trivially
Markov-regular in this context. Now the generator of such a cocycle is
$F \in B(\Comp \op \noise)$ which can be written as
\[
F = \begin{bmatrix} i \mu -\frac{1}{2}(\nu^2 +\norm{v}^2) & \langle
\nu (I-D^*D)^{1/2} w -D^* v| \\ |v\rangle & D-I \end{bmatrix}
\]
for some choice of $v \in \noise$, $\nu \in [0,\infty[$, $\mu \in
\Real$, contraction $D \in B(\noise)$ and $w \in \noise$ with
$\norm{w} \les 1$. This follows from~\eqref{con conds} (see 
also~\cite{jmLGreif}*{Theorem~5.12} or~\cite{dilation}*{Theorem~6.2}). 
It is then possible to write down the generators $G$ and $E$ of the 
positive part and partial isometry cocycles for any choice of partial 
isometry $N$ such that $N|D| = D$.

As a particular case suppose that $D$ is already a partial isometry,
let $P = |D|$, the projection onto the initial space of $D$, and take
$N = D$. Then
\begin{align*}
G &= \begin{bmatrix} -\frac{\nu^2}{2} \bigl(1 +\norm{P^\perp w}^2 
\bigr) & \langle \nu P^\perp w| \\ |\nu P^\perp w\rangle & P-I 
\end{bmatrix} \quad \text{ and } \\
E &= \begin{bmatrix} i\mu -\frac{1}{2}\norm{v}^2 -\frac{1}{2}\nu^2
\norm{P^\perp w} & \langle \nu P^\perp w -D^*v| \\ |v\rangle & D-I
\end{bmatrix}.
\end{align*}
This applies for example if we take $\noise = l^2(\Nat)$ with usual
orthonormal basis and take $D$ to be the coisometric left shift, so
that $P$ is the projection onto $\{e_1\}^\perp$. Note that if we
choose $\mu = \nu = 0$, and $v = w = 0$ then
\[
F = E = \begin{bmatrix} 0 & 0 \\ 0 & D-I \end{bmatrix}, \quad
G = \begin{bmatrix} 0 & 0 \\ 0 & P-I \end{bmatrix}.
\]
If $\Ga (Z)$ denotes the second quantisation of $Z \in B(L^2 (\Rplus;
\noise))$ then using the isomorphism $L^2 (\Rplus; \noise) \cong
L^2 (\Rplus) \ot \noise$ it follows that
\[
X_t = U_t = \Ga (M_{\ind_{[0,t[}} \ot D + M_{\ind_{[t,\infty[}} \ot 
I_\noise), \quad |X_t| = \Ga (M_{\ind_{[0,t[}} \ot P 
+M_{\ind_{[t,\infty[}} \ot I_\noise),
\]
where $M_f$ denotes multiplication by $f \in L^\infty (\Rplus)$. In 
particular since $D$ is not normal, the algebra generated by the
process $X$ is not commutative.

\medskip
On a different tack, note that equation~\eqref{L eqn} only specifies
$L$ on $\Ran |D|$, so one might be tempted to set it equal to $0$ on
the orthogonal complement, which would certainly be the case if we
replace $L$ by $L' := LN^*N$. This has the effect of apparently making
it easier to check the third of the identities in~\eqref{pi conds}
(noting that the first and second remain valid), since $L'(I-N^*N)L'^*
= 0$. But for our example above with $D$ as the left shift one finds
that $K$ now becomes
\[
K = i\mu -\tfrac{1}{2} \norm{v}^2 +\tfrac{1}{2} \nu^2 \norm{P^\perp
w}^2
\quad \Implies \quad
K +K^* +M^*M = \nu^2 \norm{P^\perp w}^2.
\]
Thus the third equation in~\eqref{pi conds} will fail for an 
appropriate choice of $\nu$ and $w$, and hence $U$ will not be partial 
isometry-valued.

\medskip
(ii) As a special case of the more general situation, suppose that $F
= \begin{sbmatrix} A & B \\ C & D-I \end{sbmatrix} \in \vN \ot
B(\khat)$ for a commutative von Neumann algebra $\vN$, and with $D$
isometric. Then $|D| = I$, $N = D$ and hence
\[
G =
\begin{bmatrix}
\frac{1}{2} (A+A^*+C^*C) & 0 \\ 0 & 0
\end{bmatrix}
\ \text{ and } \ 
E =
\begin{bmatrix}
\frac{1}{2} (A-A^*-C^*C) & -C^*D \\ C & D-I
\end{bmatrix}.
\]
In particular the positive part is $|X_t| = P_t \ot I_{\Fock_+}$ where
$P_t$ is the positive semigroup on $\init$ with generator $\frac{1}{2}
(A+A^*+C^*C)$, and all of the stochastic terms occur only in the
process $U$. Moreover in this case $\chi(E)=0$, so $U$ is an isometric 
cocycle.
\end{egs}

\bigskip
\noindent
\emph{ACKNOWLEDGEMENTS}. I am indebted to Martin Lindsay for providing
Proposition~\ref{not suff} and the example in the subsequent
remark. Many thanks to Luigi Accardi whose questions after my 
presentation of this material helped me spot an error in a previous 
version, and to Ken Duffy for facilitating the corrections.

\def\cprime{$'$} \def\polhk#1{\setbox0=\hbox{#1}{\ooalign{\hidewidth
  \lower1.5ex\hbox{`}\hidewidth\crcr\unhbox0}}}

\end{document}